\documentclass{aastex63}

\accepted{\today}
\submitjournal{ApJ}

\shorttitle{Jet and Mass Accretion Associated with RW Aur A}
\shortauthors{Takami et al.}

\usepackage{color}
\usepackage{soul,xcolor}
\setstcolor{red}

\begin{document}

\title{Possible Time Correlation Between Jet Ejection and Mass Accretion for RW Aur A\footnote{Based on observations collected at the European Southern Observatory under ESO programme 2100.C-5015.}}

\correspondingauthor{Michihiro Takami}
\email{hiro@asiaa.sinica.edu.tw}

\author{Michihiro Takami}
\affil{Institute of Astronomy and Astrophysics, Academia Sinica, 
11F of Astronomy-Mathematics Building, AS/NTU
No.1, Sec. 4, Roosevelt Rd, Taipei 10617, Taiwan, R.O.C.}

\author{Tracy L. Beck}
\affil{The Space Telescope Science Institute, 3700 San Martin Dr., Baltimore, MD 21218, USA}

\author{P. Christian Schneider}
\affil{Hamburger Sternwarte, Universit\"at Hamburg, Gojenbergsweg 112, D-21029, Hamburg, Germany}

\author{Hans Moritz G\"unther}
\affil{MIT, Kavli Institute for Astrophysics and Space Research, 77 Massachusetts Avenue, Cambridge, MA 02139, USA}

\author{Marc White}
\affil{Research School of Astronomy and Astrophysics,
College of Physical \& Mathematical Sciences,
The Australian National University,
Mount Stromlo Observatory,
Cotter Rd,
Weston Creek ACT 2611,
Australia}

\author{Konstantin Grankin}
\affil{2Crimean Astrophysical Observatory, Russian Academy of Sciences, 298409 Nauchny, Crimea}

\author{Jennifer L. Karr}
\affil{Institute of Astronomy and Astrophysics, Academia Sinica, 
11F of Astronomy-Mathematics Building, AS/NTU
No.1, Sec. 4, Roosevelt Rd, Taipei 10617, Taiwan, R.O.C.}

\author{Youichi Ohyama}
\affil{Institute of Astronomy and Astrophysics, Academia Sinica, 
11F of Astronomy-Mathematics Building, AS/NTU
No.1, Sec. 4, Roosevelt Rd, Taipei 10617, Taiwan, R.O.C.}

\author{Deirdre Coffey}
\affil{IUniversity College Dublin, School of Physics, Belfield, Dublin 4, Ireland; Dublin Institute for Advanced Studies, School of Cosmic Physics, 31 Fitzwilliam Place, Dublin 2, Ireland}

\author{Hauyu Baobab Liu}
\affil{Institute of Astronomy and Astrophysics, Academia Sinica, 
11F of Astronomy-Mathematics Building, AS/NTU
No.1, Sec. 4, Roosevelt Rd, Taipei 10617, Taiwan, R.O.C.}

\author{Roberto Galv\'an-Madrid}
\affil{
Instituto de Radioastronom\'ia y Astrof\'isica, Universidad Nacional Aut\'onoma de M\'exico, Apdo. Postal 3-72 (Xangari), 58089 Morelia, Michoac\'an, Mexico
}

\author{Chun-Fan Liu}
\affil{Institute of Astronomy and Astrophysics, Academia Sinica, 
11F of Astronomy-Mathematics Building, AS/NTU
No.1, Sec. 4, Roosevelt Rd, Taipei 10617, Taiwan, R.O.C.}

\author{Misato Fukagawa}
\affil{National Astronomical Observatory of Japan, 2-21-1 Osawa, Mitaka, Tokyo 181-8588, Japan}

\author{Nadine Manset}
\affil{Canada-France-Hawaii Telescope, 65-1238 Mamalahoa Hwy., Kamuela, HI 96743, USA}

\author{Wen-Ping Chen}
\affil{Institute of Astronomy, National Central University, Taiwan 320, Taiwan}

\author{Tae-Soo Pyo}
\affil{Subaru Telescope, 650 North Aohoku Place, Hilo, HI 96720, USA}

\author{Hsien Shang}
\affil{Institute of Astronomy and Astrophysics, Academia Sinica, 
11F of Astronomy-Mathematics Building, AS/NTU
No.1, Sec. 4, Roosevelt Rd, Taipei 10617, Taiwan, R.O.C.}

\author{Thomas P. Ray}
\affil{Dublin Institute for Advanced Studies, School of Cosmic Physics, 31 Fitzwilliam Place, Dublin 2, Ireland}

\author{Masaaki Otsuka}
\affil{Okayama Observatory, Kyoto University, Kamogata, Asakuchi, Okayama 719-0232, Japan}

\author{Mei-Yin Chou}
\affil{Institute of Astronomy and Astrophysics, Academia Sinica, 
11F of Astronomy-Mathematics Building, AS/NTU
No.1, Sec. 4, Roosevelt Rd, Taipei 10617, Taiwan, R.O.C.}

\begin{abstract}
For the active T-Taur star RW Aur A we have performed long-term ($\sim$10 yr) monitoring observations of (1) jet imaging in the  [\ion{Fe}{2}] 1.644 \micron~emission line using Gemini-NIFS and VLT-SINFONI; (2) optical high-resolution spectroscopy using CFHT-ESPaDOnS; and (3) $V$-band photometry using the CrAO 1.25-m telescope and AAVSO.
The latter two observations confirm the correlation of time variabilities between (A) the \ion{Ca}{2} 8542 \AA~ and \ion{O}{1} 7772 \AA~line profiles associated with magnetospheric accretion, and (B) optical continuum fluxes. The jet images and their proper motions show that four knot ejections occurred at the star over the past $\sim$15 years with an irregular interval of 2-6 years. The time scale and irregularity of these intervals are similar to those of the dimming events seen in the optical photometry data.
Our observations show a possible link between remarkable ($\Delta V < -1$ mag.) photometric rises and jet knot ejections. Observations over another few years may confirm or reject this trend. If confirmed, this would imply that the location of the jet launching region is very close to the star ($r$$\lesssim$0.1 au) as predicted by some jet launching models. Such a conclusion would be crucial for understanding disk evolution within a few au of the star, and therefore possible ongoing planet formation at these radii.
\end{abstract}




\section{Introduction} \label{sec:intro}

Young stellar objects of various masses and at various evolutionary stages are known to host collimated jets. Theoretical work over past decades has predicted that the jet plays an essential role for protostellar evolution, removing excess angular momentum from accreting material and allowing mass accretion to occur \citep[e.g.,][]{Blandford82,Pudritz83,Shu00,Konigl00}. This scenario has been supported by a statistical correlation between the observed mass ejection and accretion rates for many pre-main sequence stars \citep[e.g.,][]{Cabrit90,Hartigan95,Calvet97}, and observations of spinning motions in the jet \citep[e.g.,][]{Bacciotti02,Coffey04,Lee17}.
Understanding the jet driving mechanism and its detailed physical link with protostellar evolution are two of the most important issues of star formation theories.

Several theories have been proposed for the jet launching and driving, and their physical link with mass accretion. Popular magneto-centrifugal wind models have two main theories: (1) X-wind \citep{Shu00}, in which the jet launches from the inner edge of the disk ($r\ll0.1$ AU); and (2) disk wind \citep{Konigl00}, in which the jet launching region covers a larger portion of the disk surface at a few au scale. Alternative mechanisms for jet driving include magnetic pressure \citep[e.g.,][]{Machida08} and reconnection of magnetic fields between the star and the disk \citep[reconnection wind, see e.g.,][for a review]{Bouvier14}. 
Observational studies of these theories have been hampered by the limited angular resolution of present telescopes (typically as good as $\sim 0\farcs1$, corresponding to $\sim$10 au in the nearest star forming regions) \cite[see][for a review]{Frank14}. 

Simultaneous monitoring of jet ejection and mass accretion associated with pre-main sequence stars is an alternative and promising approach to test these theories. Jets exhibit knotty structures, which presumably result from episodic mass ejection with a timescale of $\sim$3 years \citep[e.g.,][]{Pyo03,Lopez03,White14a}. We can measure the epoch of each jet knot ejection by observing its position over 1-5 years and tracing it back to the origin. Accretion from the inner edge of the disk can be observed through (1) permitted line luminosities and equivalent widths; (2) redshifted absorption in permitted lines; and (3) excess UV and blue continuum emission on the stellar photosphere heated by the accretion flow \citep[see][for a review]{Calvet00}. These signatures are also time-variable \citep[see][for reviews]{Bouvier07_PPV,Bouvier14}. 
If the above mass accretion signatures immediately change when a new jet knot is ejected from the star, then the jet launching region must be associated with the stellar magnetosphere \citep{Bouvier14}. This would strongly support the X-wind or the reconnection wind models, rather than the disk wind model, in which the jet is launched from the disk surface at radii up to 2-3 au. 

We have been conducting such monitoring observations from 2010 for three of the best-studied T-Tauri stars  (RW Aur A, RY Tau, DG Tau). This paper highlights our observations of RW Aur A to date. 
This star is one of the first young stars that drew researchers' particular attention due to its peculiar optical emission line spectra \citep[e.g.,][]{Herbig45,Appenzeller82} and variability \citep[e.g.,][]{Herbig48,Gahm70}. These early studies were followed by a number of spectroscopic observations of these emission lines in order to understand magnetospheric accretion and wind activities close to the star \citep[e.g.,][]{Petrov01a,Alencar05,Takami16,Facchini16}. While RW Aur A is associated with a resolved companion 1\farcs5 away \citep[RW Aur B, e.g.,][]{Joy44,Reipurth93,White01,Bisikalo12}, a few spectroscopic studies suggest that the star is also associated with a spectroscopic binary \citep[e.g.,][]{Gahm99,Petrov01a}. The star appears to have been photometrically stable over many years \citep[e.g.,][]{Beck01,Grankin07}, however, it has shown peculiar photometric changes at variety of wavelengths since 2010 \citep[e.g.,][]{Rodriguez13,Rodriguez18,Schneider15,Petrov15,Shenavrin15,Bozhinova16,Lamzin17,Gunther18}. Table \ref{tbl:rwaur} summarizes key stellar parameters for RW Aur A.
%

RW Aur A is associated with a bipolar asymmetric jet, consisting of a brighter redshifted jet and a fainter blueshifted counterpart \citep[][]{Mundt98,Hirth94,Hirth97,Bacciotti96,Berdnikov17}. 
Extensive observations at high-angular resolutions ($\sim$0\farcs1) have been made to study their structure, excitation and kinematics close to the star \citep[e.g.,][]{Dougados00,Woitas02,Pyo06,Beck08,Coffey08,Hartigan09} and their spinning motions \citep{Coffey04,Coffey12}. 
The asymmetry in jet emission is either due to different mass ejection rates between the redshifted and blueshifted jets \citep{Liu12}, or the different conditions of surrounding gas on the two sides but with similar mass ejection rates \citep{Melnikov09}.

The rest of the paper is organized as follows. In Section \ref{sec:obs} we describe our
observations of the jet knot ejections, optical line profiles and optical continuum fluxes. In Section \ref{sec:results} we highlight the results of these observations. In Section 4 we discuss possible implications for the mechanism of jet ejection and a physical link with mass accretion.

\begin{table}
\caption{Stellar Properties of RW Aur A \label{tbl:rwaur}}
\begin{tabular}{ll}
\tableline\tableline
Distance				& 162$\pm$2 pc\tablenotemark{a}	\\
Mass				& 1.3	$\pm$0.2 $M_\sun$\tablenotemark{b}	\\
Spectral Type			& K4-K5\tablenotemark{c} 		\\
Stellar Luminosity		& 1.7 $L_\sun$\tablenotemark{b}	\\
Age					& 8.3 Myr\tablenotemark{b}	\\
Mass Accretion Rate		& $3\times 10^{-8} M_\sun$ yr$^{-1}$ \tablenotemark{b}		\\
\tableline\tableline
\end{tabular}
\tablenotetext{a}{Gaia DR2 \citep{GaiaDR2}. The measurement of RW Aur A has a large uncertainty, hence we adopt that for the companion star RW Aur B 1\farcs5 away from the primary star.
\tablenotetext{b}{\citet{White01}}
\tablenotetext{c}{For the bright stable periods. The stellar absorption lines were not clearly observed during the dimming periods in 2010 and 2014 \citep{Takami16}.}
}
\end{table}


\section{Observations and Data Reduction} \label{sec:obs}


\subsection{Jet Imaging} \label{sec:obs:jet}
Integral field spectroscopic observations of the [\ion{Fe}{2}] 1.644 $\micron$ line were obtained using NIFS at the Gemini North Telescope and SINFONI at the Very Large Telescope. 
The $H$ grating for these instruments yielded a spectral resolution $R$$\sim$5500 ($\Delta v \sim 55$ km s$^{-1}$)
and $\sim$3000 ($\Delta v \sim 100$ km s$^{-1}$), respectively, at 1.5-1.8 $\micron$, over a field of view (FOV) of approximately 3\arcsec$\times$3\arcsec.
Table \ref{tab:IFUs} summarizes our observations to date. The star was placed at the center of the FOV for some epochs, and placed near the edge of the FOV for the others to cover the redshifted jet (i.e., the brighter jet) with a large spatial area. The point-spread function (PSF) of the adaptive optics observations, which we measured using the target star, consists of core and halo components, and we fit them using two separate gaussians. The FWHM of the core component indicates an angular resolution of the observations of 0\farcs10-0\farcs16. We measured the core-to-total flux ratio of 0.3--0.6. 
An occulting mask with a 0\farcs2 diameter was used to block stellar light in the observations on 2012 October 20. For this date we also obtained short exposures without a coronagraphic mask and used them to measure the PSF. See Table \ref{tab:IFUs} for details of these parameters.

\begin{deluxetable*}{cllcrrcl}
\tablecaption{Log of the NIFS and SINFONI observations\label{tab:IFUs}}
\tablewidth{0pt}
\tablehead{
\colhead{Date} &
\colhead{Instrument} &
\colhead{Observing} &
\colhead{Photometric} &
\colhead{$t_{exp}$} &
\colhead{$n_{exp}$} &
\colhead{Core FWHM\tablenotemark{b}} &
\colhead{$f_{core}$\tablenotemark{a,b}}
\\
\colhead{YYYY-MM-DD} & 
& 
\colhead{run ID} & 
& 
\colhead{(s)} & 
& 
\colhead{(arcsec)} & 
}
\startdata
2012-10-20 	& NIFS		& GN-2012B-Q-99 & $\circ$	& 600	& 9		& 0.16	& 0.61	\\
2014-02-28	& NIFS		& GN-2014A-Q-29 & $\circ$	& 60		& 12		& 0.15	& 0.48	\\
2014-12-29	& NIFS		& GN-2014B-Q-18 & $\circ$	& 84		& 20		& 0.15	& 0.38	\\
2017-02-15	& NIFS		& GN-2017A-FT-1 & $\circ$	& 55		& 36		& 0.15	& 0.49	\\
2017-12-08	& SINFONI	& 2100.C-5015(A) & 		& 4		& 140	& 0.12	& 0.28	\\
2017-12-11	& SINFONI	& 2100.C-5015(A) & 		& 4		& 140	& 0.10	& 0.35	\\
2018-08-21	& NIFS		& GN-2018B-Q-141 & 		& 55		& 17		& 0.14	& 0.50 	\\
2018-08-31	& NIFS		& GN-2018B-Q-141 & $\circ$	& 55		& 3		& 0.12	& 0.39	\\
2018-09-16	& NIFS		& GN-2018B-Q-141 & 		& 55		& 19		& 0.12	& 0.50	\\
2019-10-07	& NIFS		& GN-2019B-Q-132 & 		& 55		& 36		& 0.16	& 0.53	\\
\enddata
\tablenotetext{a}{Fractional flux of the PSF core (see text.)}
\tablenotetext{b}{Measured at 1.65 $\micron$.}
\end{deluxetable*}

Data reduction was made using the Gemini IRAF package, pipelines provided by European Southern Observatory, and software we developed using PyRAF, numpy, scipy, and astropy on python. For NIFS data, we used the Gemini IRAF package for sky subtraction, flat-fielding, the first stage of bad pixel removals, 2 to 3 dimensional transformation of the spectral data, and wavelength calibration. We then used our own software for stacking data cubes for each date, telluric correction, flux calibration, extraction of the cube for the target emission line, additional removal of bad pixels, and continuum subtraction. We have also corrected a flux loss with the PSF halo, as the jet structures we are interested in are significantly smaller than the PSF halo ($>$0\farcs5).
We have used identical processes for the SINFONI data but data stacking was made using the observatory pipeline.

Table \ref{tab:IFUs} also shows whether the observations for each date were made during photometric conditions. While one would regard the absolute flux as reliable only at such conditions, the less accurate calibration for the remaining dates does not affect the discussion and conclusions in later sections.

We found a marginal error (1$\degr$-2$\degr$) in the actual image position angle from those set for the NIFS and SINFONI observations. This was corrected by measuring the position angle (PA) toward the binary companion RW Aur B ($d \sim$ 1\farcs5), adopting PA=$254\fdg47\pm0\fdg03$ based on the GAIA DR2 measurements in 2014-2016. We did not correct the effect of binary motion as it is minimal: the binary PA was $255\fdg46\pm0\fdg07$ in 1994 \citep{White01}, yielding a change of binary PA of $\sim 0\fdg05$ yr$^{-1}$.
We also found a systematic error in wavelength calibration for the pipelined SINFONI data of $\sim$60 km s$^{-1}$, by comparing them with those of modeled telluric atmospheres \citep[ATRAN, ][]{ATRAN}. 
This will not affect the discussion and conclusions of this paper (see Section \ref{sec:results}). 

The data of late 2017 and mid-2018 were obtained on a few dates. The intensity distribution of the [\ion{Fe}{2}] line was consistent between the visits in each season.  We averaged these cubes, adjusting their weights to maximize the signal-to-noise. For the mid-2018 data we then corrected the flux based on the observations made during photometric conditions.


\subsection{Optical Line Profiles}  \label{sec:obs:CFHT}

Optical high-resolution spectroscopy was made using the 3.6-m Canada-France-Hawaii Telescope (CFHT) with ESPaDOnS, covering the wavelength range 3700--10500 \AA. The spectra were obtained using  the ``object+sky spectroscopy mode" with a spectral resolution of 68,000.
Table \ref{tbl:cfht} shows the log of the observations for nine autumn-winter semesters (August-January). The observations in each semester were made with 2--4 observing runs, and 1--7 visits during each run, with intervals of 1--10 nights.
Seeing was 0\farcs8 or less for about 80 \% of the visits, and it exceeded 1\arcsec~ during $\sim$ 10 visits, reaching up to 1\farcs5. 
A single 360-s exposure was made for most of the nights.
More exposures were obtained for a few nights to increase signal-to-noise in cloudy conditions.
Data were reduced using the standard pipeline ``Upena'' provided by CFHT.
The spectra obtained on the same nights were nearly identical to each other, hence we obtained a weighted-averaged spectrum to represent the line profiles for these dates.
Some data obtained by early 2015 have already been published in \citet{Chou13} and \citet{Takami16}. See \citet{Takami16} for other details of the observations, data reduction and calibration. 

\begin{table}
\caption{Log of the spectroscopic observations \label{tbl:cfht}}
\begin{tabular}{lcl}
\tableline\tableline
Semester\tablenotemark{a} & Run & Dates (YYYY-MM-DD)\\ \tableline
2010B	& 1	&	2010-10-(16, 21)	\\
		& 2	&	2010-11-(17, 21, 25, 27)	\\

2011B	& 1	&	2011-08-20		\\
		& 2	&	2012-01-(05, 09, 11, 15)	\\

2012B	& 1	&	2012-09-(26, 29)	\\
		& 2	& 	2012-11-(25, 28) ; 2012-12-(01, 08)	\\
		& 3	&	2012-12-(22, 25, 28) \\

2013B	& 1	&	2013-08-(15, 17, 21, 28) \\
		& 2 	&	2013-09-26 \\
		& 3	&	2013-11-23 \\
		& 4	&	2014-01-(10, 11, 15, 16, 17, 19, 20) \\

2014B	& 1	&	2014-08-(15, 19) \\
		& 2	&	2014-09-(04, 10, 15) \\
		& 3	&	2014-11-(05, 08) \\
		& 4	&	2014-12-(20, 22, 29) ; 2015-01-(07, 11) \\		

2015B	& 1	&	2015-09-(23, 25) ; 2015-10-(01, 02) \\
		& 2	&	2015-10-30 \\
		& 3	&	2015-11-(25, 27, 30) ; 2015-12-02 \\
		& 4	&	2016-01-(14, 16, 23, 25) \\		

2016B	& 1	&	2016-08-(04, 08, 11, 14) \\
		& 2	&	2016-10-(12, 15, 17, 20) \\
		& 3	&	2016-12-14 \\
		& 4	&	2017-01-(18, 20, 22) \\		

2017B	& 1	&	2017-09-(07, 10) \\
		& 2	&	2017-11-(01, 03, 05, 08) \\
		& 3	&	2017-12-(28, 30) ; 2018-01-(02, 03, 05, 06, 07, 10) \\

2018B	& 1	&	2018-08-(17, 22) \\
		& 2	&	2018-09-30 ; 2018-10-02 \\
		& 3	&	2018-10-(21, 23) \\
		& 4	&	2018-11-(16, 20) \\		
		& 5	&	2018-12-(22, 24, 26) \\		
\tableline
\end{tabular} \\
\tablenotetext{a}{from August to January the following year.}
\end{table}

For this paper we present the line profiles for \ion{Ca}{2} 8542 \AA~and  \ion{O}{1} 7772 \AA, for which \citet{Takami16} show clear time variabilities potentially related to the jet ejections. We removed adjacent photospheric lines using a spectrum of the weak-lined T Tauri star Par 1379 (K4) as for \citet{Takami16}. These line profiles do not clearly show evidence for contaminating emission from RW Aur B, which has a remarkably different spectrum from RW Aur A, even at the largest seeing (1\farcs0--1\farcs5). The signal-to-noise of the adjacent continuum was $\ga$50 per resolution element for the bright photometric states, but significantly lower for some spectra obtained in the faint states (see Section \ref{sec:results} for the photometric variability during the observations). We therefore convolved the line profiles using a gaussian to increase the signal-to-noise ratio. The actual velocity resolutions of the  \ion{Ca}{2} and  \ion{O}{1} profiles shown in later sections are 10 and 30 km s$^{-1}$, respectively.


\subsection{Optical Continuum Fluxes}  \label{sec:obs:phot}

Photometric observations were made using the Crimean Astrophysical Observatory (CrAO) 1.25-m telescope (AZT-11) with the Finnish five-channel photometer and the ProLine PL23042 CCD detector \citep{Petrov15} and an entrance diaphragm of 15\arcsec. We will use data for 300 visits of the $V$-band data obtained between 2009 to 2018. 
We added data from the American Association of Variable  Star Observers (AAVSO) archive \citep{aavso}, downloading the data for 4,725 visits between mid-2005 and early 2019. These observations include the flux from the companion RW Aur B \citep[12.9-13.6 mag. in $V$-band;][see also Section \ref{sec:obs:jet}]{White01,Antipin15}. Comparisons with resolved photometry and spectroscopy indicate that the optical time variabilities of the RW Aur A+B system are primarily due to RW Aur A \citep[e.g.,][Section 3]{Antipin15,Schneider15,Gunther18}.



\section{Results} \label{sec:results}

\begin{figure}[ht!]
\includegraphics[width=9cm]{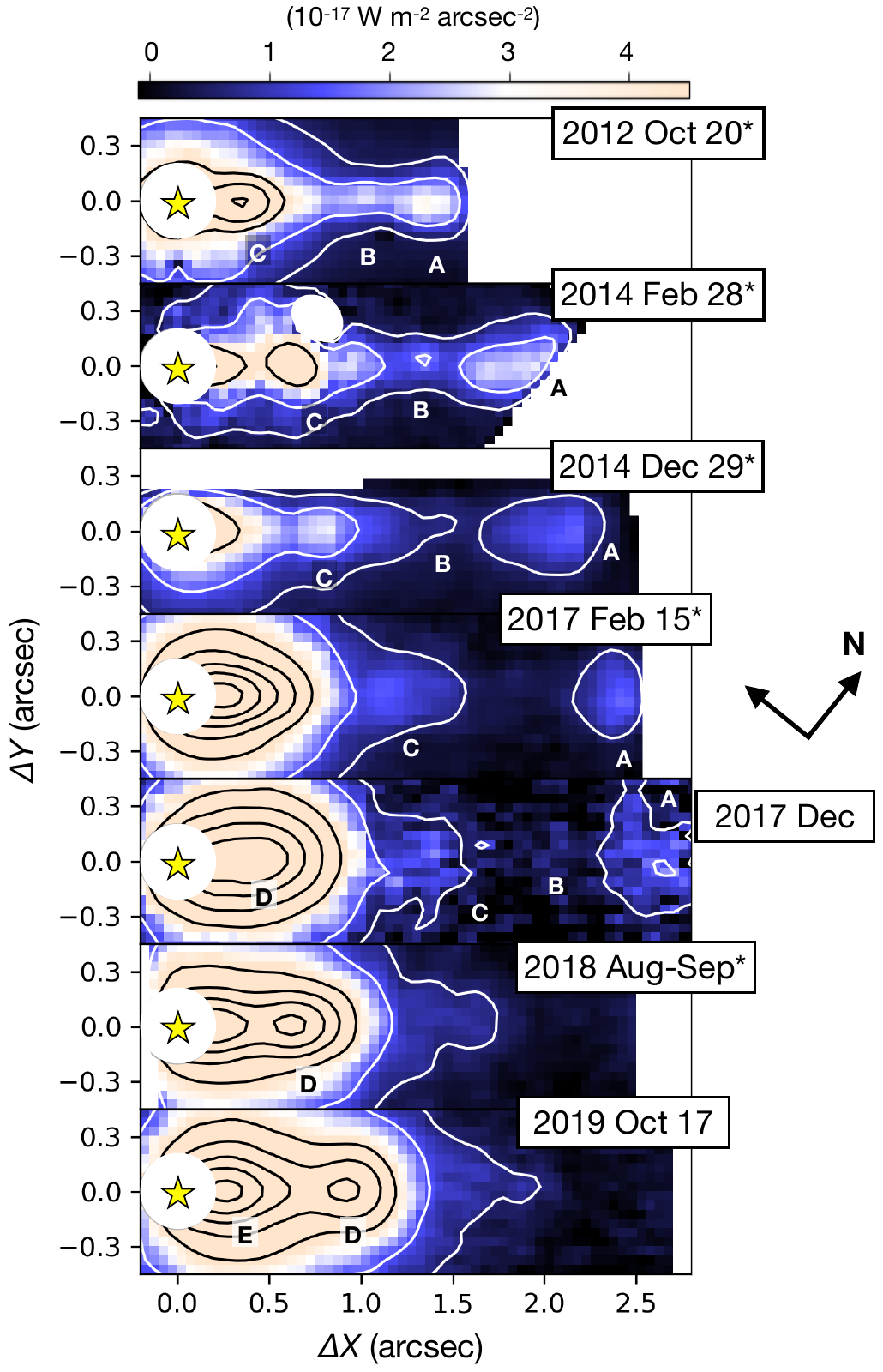}
\caption{The velocity-integrated images 
of the [\ion{Fe}{2}] 1.644 \micron~emission associated with the redshifted jet from RW Aur A.
The spatial offsets $\Delta X$ and $\Delta Y$, along and across the jet axis (PA=309\arcdeg), respectively, are shown with regard to the stellar position.
The intensity distribution within 0\farcs2 of the star is masked because of imperfect continuum subtraction.
The contour levels are arbitrarily chosen to try to clearly show the jet structures.
The identified knots are labeled as A-E. 
%
%
In the integration map for 2014 Feb 28, the emission at ($\Delta X$,$\Delta Y$) = (0\farcs8,0\farcs3) is masked due to imperfect subtraction of the continuum source in the sky frames.
The asterisk next to the date indicates the data for which absolute intensity is highly reliable (Section \ref{sec:obs:jet}).
Different spatial coverages of the jet at different epochs result from different image PAs and different locations of the star in the FOV.
\label{fig:jet}}
\end{figure}

Figure \ref{fig:jet} shows the images
of the redshifted jet of RW Aur for seven epochs. 
For the NIFS data we integrated each data cube over $V_{Hel}$=50 to 200 km s$^{-1}$ to obtain the integrated maps of the redshifted jet, and a spatial range $\Delta Y$ of ($-$0\farcs15, 0\farcs15) from the star across the jet axis (PA=309\arcdeg) to obtain the PV diagrams. These velocity and spatial ranges cover most of the observed line emission. 
For the SINFONI data, which have a systematic error for velocity calibration (Section \ref{sec:obs}), we integrated the data cube over $\Delta v = 200$ km s$^{-1}$ to cover most of the jet emission and obtain the integrated map.

In the figure we identify five knots labeled as A-E. As expected from previous observations of the RW Aur jet \citep[e.g.,][]{Lopez03}, we observe larger offsets for newer epochs because of the proper motions. Table \ref{tab:knots} shows the positions of the Knots B-E for individual epochs. To measure their positions, we first integrated each data cube over the above velocity and spatial ($\Delta Y$) ranges for the integrated maps and PV diagrams, and obtained one-dimensional intensity distribution along the jet axis. We then applied a polynomial fitting using 4-6 pixels near the peak, and measured the peak position. 

\begin{figure}[ht!]
\includegraphics[width=8cm]{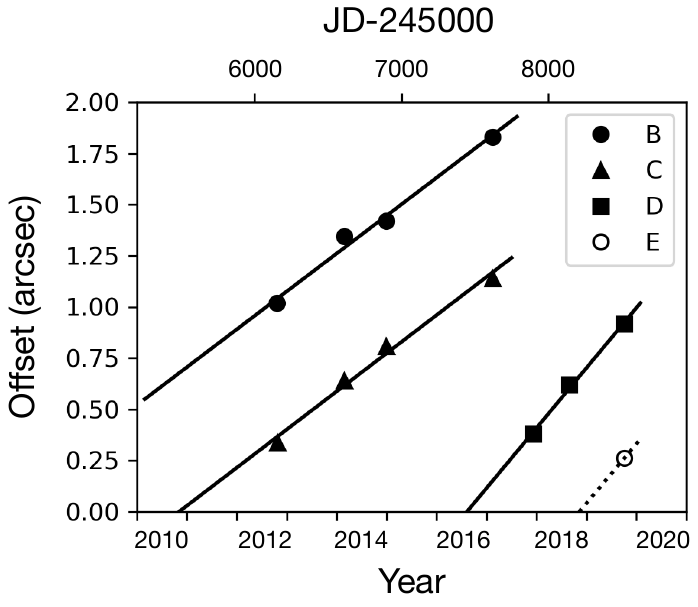}
\caption{Linear fit of the knot positions. For Knot E we draw a line assuming a proper motion of 0\farcs29 yr$^{-1}$.
\label{fig:motion}}
\end{figure}

\begin{deluxetable*}{ccccccccccch}
\tabletypesize{\scriptsize}
\tablecaption{Knot Positions and Proper Motions\label{tab:knots}}
\tablewidth{0pt}
\tablehead{
\colhead{Knot} &
\multicolumn{7}{c}{Offset (arcsec) on YYYY-MM\tablenotemark{a}} &
\colhead{Proper Motion} &
\colhead{Fitting Error} &
\colhead{Origin} &
\nocolhead{$V_{Hel}$}
\\
&
\colhead{2012-10} & 
\colhead{2014-02} & 
\colhead{2014-12} & 
\colhead{2017-02} & 
\colhead{2017-12} & 
\colhead{2018-08/09} & 
\colhead{2019-10} & 
\colhead{(arcsec yr$^{-1}$)} &
\colhead{(arcsec)} &
\colhead{JD-2450000} &
\nocolhead{(km s$^{-1}$)}
}
\startdata
B	& 1.018	& 1.345	& 1.420	& 1.829	& ---		& ---		& ---		& 0.185$\pm$0.019	& 0.041	& 4161$\pm$284	& 101$\pm$8	\\
C	& 0.339	& 0.640	& 0.808	& 1.140	& --- 		& 1.624	& --- 		& 0.186$\pm$0.014	& 0.034	& 5506$\pm$112	& 103$\pm$5	\\
D	& --- 		& --- 		& ---		& ---		& 0.395	& 0.620	& 0.919	& 0.293$\pm$0.016	& 0.014	& 7622$\pm$44	& 136$\pm$6	\\
E	& ---		& ---		& ---		& ---		& ---		& ---		& 0.263	& ---				& ---		
		& (8432)\tablenotemark{b}			
		& 131	\\
\enddata
\tablenotetext{a}{See Table \ref{tab:IFUs} for the exact dates of the observations.}
\tablenotetext{b}{Assuming a proper motion of 0\farcs29 yr$^{-1}$.}
\end{deluxetable*}

For Knots B-D we applied a linear fit to these positions to derive the proper motion and the date of ejection at the star (Figure \ref{fig:motion}, Table \ref{tab:knots}).
For each knot we derived 1-$\sigma$ uncertainties for these parameters using scipy.optimize.curve\_fit.
The proper motions of 0\farcs19-0\farcs29 yr$^{-1}$ are similar to those of the knots ejected in 1980-1997 \citep[0\farcs16-0\farcs26;][]{Lopez03}.
%
%
The 1-$\sigma$ uncertainties for the dates of ejection for Knots B, C, D are $\sim$280, $\sim$100, and $\sim$40 days, respectively.
The large uncertainty for Knot B primarily results from the fact that the measurements were made long after it was ejected from the star (5--10 years). The uncertainty for Knot D is better than that for C, perhaps due to its brighter nature.
 For Knot E we tentatively assume a proper motion nearly identical to Knot D (0\farcs29 yr$^{-1}$) to estimate an approximate date of mass ejection from the star.
We skip the above analysis for Knot A because of its relatively blurred structure and large offsets from the star, which cause a large uncertainty in the analysis below. 

Table \ref{tab:knots} shows that jet knot ejections over the past $\sim$15 years have occurred with an irregular interval of 2-6 years. A similar trend was also observed in previous jet ejections from RW Aur A \citep{Lopez03} and another active pre-main sequence star, DG Tau \citep[][]{Pyo03,Agra11,White14a}. 

\begin{figure*}[ht!]
\includegraphics[width=18cm]{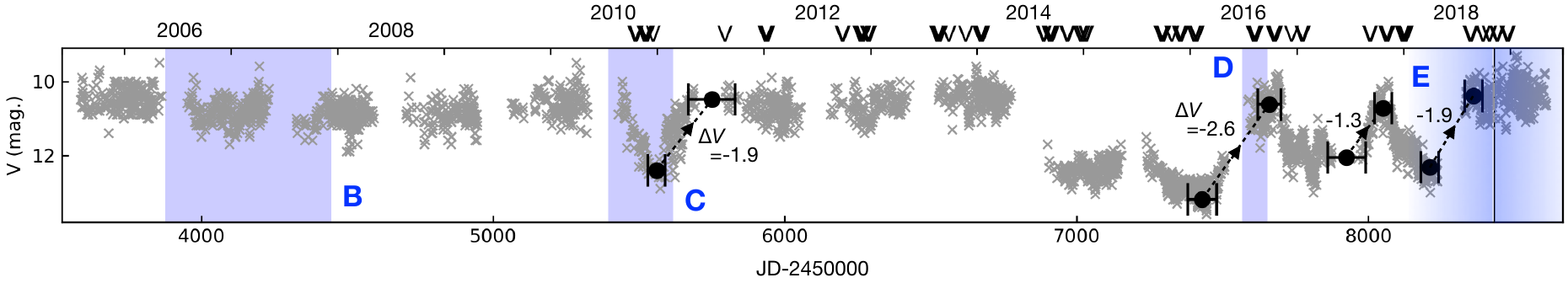}
\caption{$V$-magnitude of the RW Aur AB system from the middle 2005 to early 2019. The blue boxes B-D indicate the dates of the knot ejection from the star tabulated in Table \ref{tab:knots} with 1-$\sigma$ uncertainties. A tentative date of the ejection for Knot E is also shown in blue but with blurred boundaries. The `V' marks at the top of the box indicate the dates of the spectroscopic observations (Section \ref{sec:obs:CFHT}).
The figure shows remarkable ($\Delta V < -1$) photometric rises in 2011, 2016, 2017, and 2018. Before and after each rise, we measure a median magnitude at a range indicated by the horizontal bar, plot it using a large dot, and provide the change in $V$-band magnitude.
\label{fig:Vmag}}
\end{figure*}

Figure \ref{fig:Vmag} shows the $V$-band magnitude of the RW Aur AB system between mid-2005 and early 2019. The results by early 2018 have been published in \citet{Rodriguez13,Rodriguez18,Petrov15,Gunther18,Dodin19}.
In Figure \ref{fig:Vmag}
we overplot (1) the date of ejection of Knots B-E from the star with uncertainties; and (2) the dates of our spectroscopic observations. The figure shows that Knot C appears to have been ejected from the star during the dimming state in 2010, or the subsequent photometric rise located at a 1-$\sigma$ level. Knot D appears to have been ejected near the end of the photometric rise in 2016. Knot E may also have been ejected during or at the end of the photometric rise, but measurements of the proper motions over another few years are required to confirm or reject this trend. Figure \ref{fig:Vmag} does not show any photometric rise associated with the ejection of Knot B. To further discuss a possible link between the jet knot ejection and the photometric rises, we measured the magnitude before and after the photometric rises at/near the ejection of Knots CDE, and another remarkable ($\Delta V < -1$) rise in 2017. These results are also shown in Figure \ref{fig:Vmag} .


\begin{figure}[ht!]
\includegraphics[width=18cm]{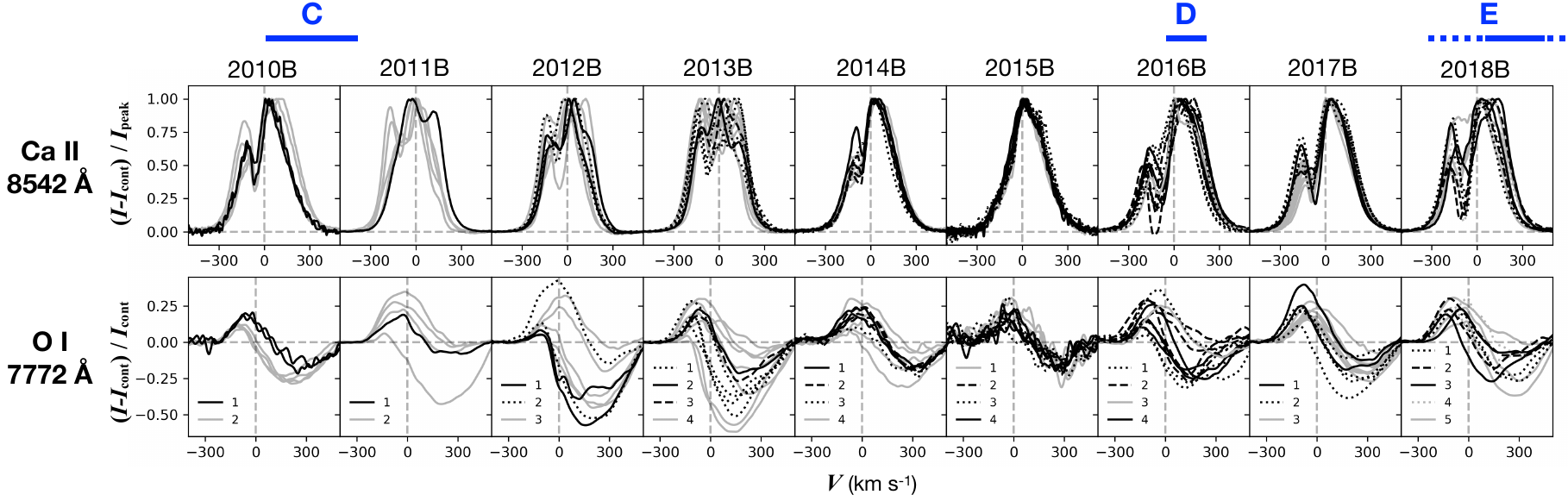}
\caption{The \ion{Ca}{2} 8542 \AA~ and \ion{O}{1} 7772 \AA~ line profiles observed during the 2010B-2018B semesters. 
The profiles with different colors (black/gray) styles (solid/dashed/dotted) were observed on different observing runs (see Table \ref{tbl:cfht}). The \ion{O}{1} profiles are normalized to the continuum flux, while the  \ion{Ca}{2} profiles are normalized to the peak flux to clarify the variabilities discussed in the text. All the \ion{O}{1} 7772 \AA~profiles observed in the 2015B semester and a few in the 2010B semester have a relatively low signal-to-noise (see Section 2). At the top of the figure we approximately indicate when Knots C-E were ejected.
\label{fig:profs:semesters}}
\end{figure}

\begin{figure}[ht!]
\includegraphics[width=18cm]{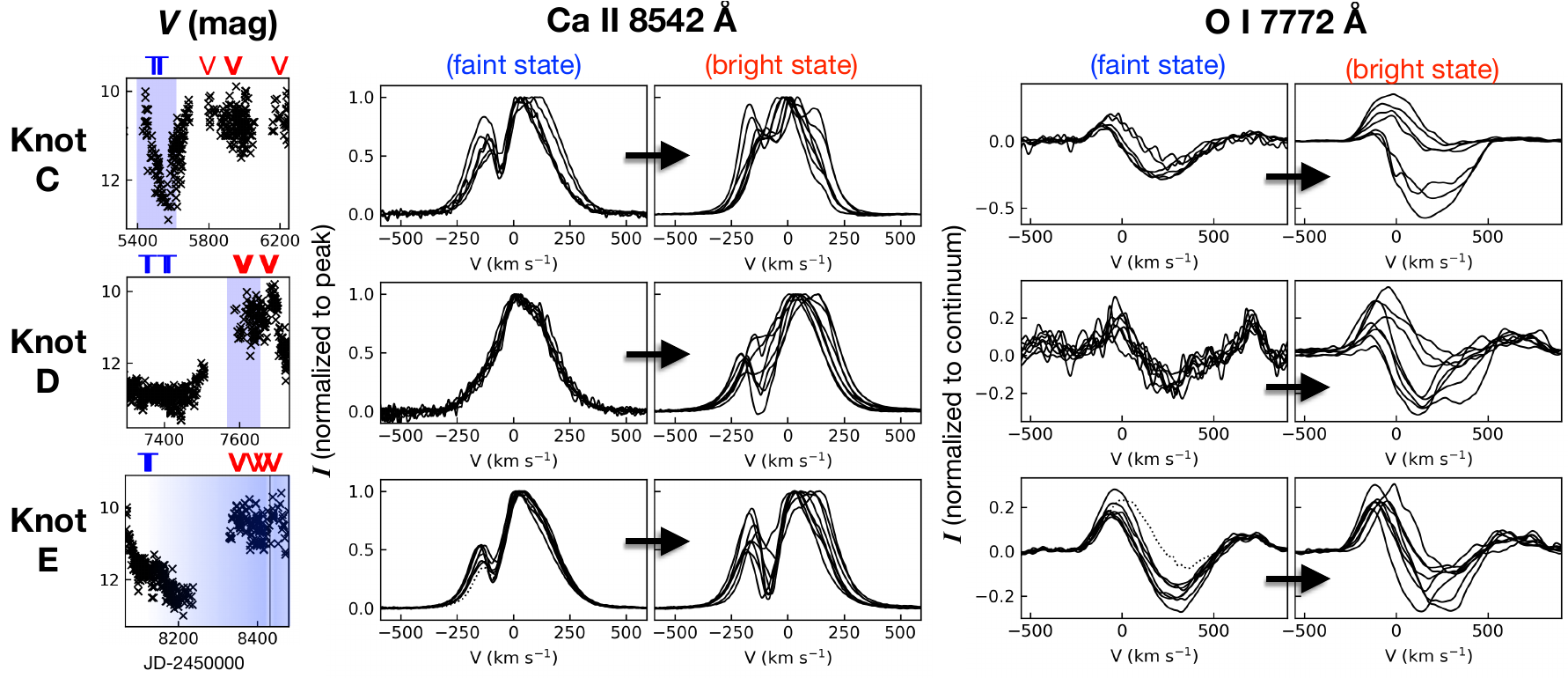}
\caption{($left$) Same as Figure \ref{fig:Vmag} but near the ejection of Knots C-E from the star. The blue `T' and red `V' marks at the top of each box indicate the dates of the CFHT observations at the faint dimming states and the bright stable states, respectively. 
($middle$, $right$) The \ion{Ca}{2} 8542 \AA~and \ion{O}{1} 7772 \AA~line profiles for faint and bright periods near the ejections of Knots C-E from the star. As for Figure \ref{fig:profs:semesters}, the \ion{O}{1} profiles are normalized to the continuum flux, while the \ion{Ca}{2} profiles are normalized to the peak flux. Some profiles show small fluctuations in velocity ($\Delta v$$\sim$10 and 30 km s$^{-1}$ for \ion{Ca}{2} and \ion{O}{1}, respectively) due to their relatively low signal-to-noise. The \ion{O}{1} profile observed on 2018 January 7 (dotted line) shows a relatively large deviation from the other profiles in the panel of the faint state for Knot E. We use dotted curves for the line profiles observed on the former date, both for \ion{Ca}{2} and  \ion{O}{1}.
\label{fig:profs:knots}}
\end{figure}

Figure \ref{fig:profs:semesters} shows the \ion{Ca}{2} 8542 \AA~and \ion{O}{1} 7772 \AA~line profiles. During the 2011B-2013B semesters, when the system was bright and photometrically stable, the \ion{Ca}{2} profiles show complicated variabilities near the peak, while redshifted \ion{O}{1} absorption shows a large variation \citep{Takami16}. These line profiles are more stable in 2010B, 2014B and 2015B, i.e., when the star became fainter in $V$-band.
Such a trend is less clear for 2016B-2018B in Figure \ref{fig:profs:semesters} because of the complex photometric variabilities during this period.
Figure \ref{fig:profs:knots} shows line profiles before and after the photometric rises associated with Knots C-E. As for 2010B-2015B in Figure \ref{fig:profs:semesters}, the line profiles are relatively stable during the faint periods (i.e., before the photometric rise) but these show complex/large variabilities in the bright periods (i.e., after the photometric rise). For the \ion{O}{1} profiles near the ejection of Knot E, this trend is clear only if we ignore the dotted profile observed on 2018 Jan 7. It is not clear what causes the deviation of this profile from the others. We believe that the peculiarity of the line profile on this specific date does not significantly affect the discussion and conclusions below.

\section{Discussion}  \label{sec:discussion}

In this section we discuss possible implications for the mechanism of jet ejection and a physical link with mass accretion. The variabilities in optical continuum and line fluxes could be associated with mass accretion in general (see Section \ref{sec:intro}), however, RW Aur A is known for the complicated nature of its variability. In Section \ref{sec:discussion1} we summarize the present understanding of these variabilities. In Section \ref{sec:discussion2} we discuss their possible link with the jet knot ejections.

\subsection{Origin of Variability of Optical Flux and Spectra} \label{sec:discussion1}

The blue excess continuum and optical permitted line profiles associated with pre-main sequence stars are due to mass accretion from the inner edge of the disk to the star \citep[e.g.,][for reviews]{Calvet00,Najita00}. According to the current paradigm,  the stellar magnetosphere is associated with the inner edges of the circumsteller disk, and they regulate the stellar rotation. Mass accretion from the disk to the star occurs through this magnetic field. These accretion flows, along so-called magnetospheric accretion columns, are associated with optical and near-infrared permitted line emission (including \ion{Ca}{2} and \ion{O}{1}), in particular with the broad component (a full width half maximum velocity $V_{\rm FWHM} > 100$ km s$^{-1}$). Accretion shocks at the stellar photosphere cause hot spots, with a typical temperature of $\sim 10^4$ K \citep[see][for the measurement of temperatures]{Gullbring98}, and these add blue excess continuum to the photospheric emission ($T_{\rm eff} \sim 4000$ K).
The luminosities of the blue excess continuum and permitted lines monotonically increase with the mass accretion rate. The accretion shocks may also induce X-ray radiation \citep[e.g.,][]{Lamzin99}. All of the above emission is associated with the surface of the star and regions very close to the star ($r \ll 0.1$ AU).

Throughout, the observed optical continuum flux can vary with the mass accretion rate.
In addition, the optical flux can also change with obscuration by dusty blobs or a wind crossing in front of the star, or by structures on an optically thick dusty disk. A group of such young stars have been traditionally classified as UX Ori-type variables \citep[e.g.,][]{Herbst94}. To date, 
many authors support a scenario of dust obscurations to explain decreases of optical fluxes toward RW Aur A based on photometry at a variety of wavelengths \citep[][]{Rodriguez13,Rodriguez18,Schneider15,Petrov15,Shenavrin15,Lamzin17,Gunther18}, spectroscopy \citep{Petrov15,Facchini16,Koutoulaki19} and polarimetry \citep{Dodin19}.

\citet{Takami16} discussed some reservations of the occultation scenario for RW Aur A, and also the possibility 
of adding the mechanism of time variable mass accretion to explain the variabilities of optical continuum flux and line profiles.
We update the discussion adding recent work by other groups.


\subsubsection{Occultation Scenario} \label{sec:discussion11}

This scenario explains a variety of observations, including the color change in the optical/X-ray continuum \citep[][]{Petrov15,Schneider15,Gunther18,Dodin19} and optical polarization \citep[][]{Dodin19}. Near-IR and X-ray observations by \citet{Shenavrin15,Schneider15} show the presence of hot dust and gas components, respectively, associated with obscuring material close to the star.

However, the occultation scenario cannot simply explain the spectral variations shown in Section \ref{sec:results}. The spectra and line profiles should not change if the star, the accretion flows and a wind are uniformly occulted. The hot spots on the stellar surface, the emission from accretion flows and a wind are not uniform, therefore the spectral variations would occur if (1) the occulter allows only a part of the stellar surface or the inflow/outflow to be observed; or (2) while the direct fluxes from the star and the inflow/outflow are fully occulted, some emission is still observed via scattering from circumstellar dust. However, it is not clear if these can explain the similar H$\alpha$ profiles and the H$\alpha$, \ion{Ca}{2}, \ion{O}{1} and \ion{He}{1} equivalent widths through the entire period of observations \citep{Takami16}.



\subsubsection{Accretion Scenario} \label{sec:discussion12}

The bright photometric periods would be due to high mass accretion rates, which cause a relatively bright blue excess continuum. The high accretion rates would simultaneously induce a magnetic Rayleigh-Taylor (RT) instability in the accretion flows  \citep[e.g.,][]{Romanova08,Kurosawa13}, and as a result, yield a complicated time variation in the \ion{Ca}{+2} line profiles and a large variation of redshifted absorption in \ion{O}{+1} and \ion{He}{+1} lines. This scenario could naturally explain the correlation of time variabilities between the optical continuum flux and line profiles. 

However, this scenario cannot explain the time variability of optical polarization, which increases remarkably during the faint period \citep{Dodin19}.  A combination of a reflection nebula and obscuration of the stellar light is still required to explain the optical polarization even with this scenario. X-ray observations by \citet{Schneider15,Gunther18} indicate relatively large sizes for the grains in the obscuring material, and also an enhancement of the Fe abundance in hot gas. \citet{Gunther18} pointed out that these trends could result from the breakup of planetesimals in the accretion flow. \citet{Garate18} executed numerical simulations and demonstrated that an enhanced disk accretion rate would alter the physical conditions of the inner disk, and enhance large grains close to the star as a result. A careful investigation is necessary to determine whether this scenario can also explain the relatively stable near-IR CO spectra associated with the surface of the inner disk \citep{Koutoulaki19}.


\subsubsection{Caveats on Both Scenarii} \label{sec:discussion12}

\citet{Takami16} found the absence of optical photospheric absorption and redshifted absorption in the \ion{Li}{1} 6708 \AA~line in the high-resolution spectra observed in the faint periods. Neither scenario described above can simply explain these spectral changes from the bright periods.


\subsection{Physical Nature of Jet Ejection} \label{sec:discussion2}

RW Aur A appears to have started exhibiting photometric dimming events in 2010 \citep[][Section \ref{sec:results}]{Rodriguez13,Rodriguez18,Petrov15,Gunther18,Dodin19}. Figure \ref{fig:Vmag} shows
remarkable dimming events ($\Delta V > 1$) in late 2010 to early 2011; mid-2014 to mid-2016; late 2016 to late 2017; and late 2017 to mid-2018, with an irregular time interval of 1--4 years. 
Jet knot ejections for the past $\sim$15 years have occurred with an irregular interval with a similar time scale (2-6 year), as shown in Section \ref{sec:results}.

In particular, the ejections of Knots C-E occurred at/near a remarkable photometric rise ($\Delta V < -1$) within uncertainties of the measurements, indicating their possible link. Such a link would be explained if (1) the time variations of the optical continuum flux and line profiles shown in Section \ref{sec:results} are associated with time variable mass accretion, as discussed in Section \ref{sec:discussion12}; and (2) there is a physical link between jet ejection and mass accretion (Section \ref{sec:intro}). These may not be surprising because the mass ejection rate estimated using optical forbidden lines and the mass accretion rate inferred from the blue excess continuum are statistically correlated among many pre-main sequence stars \citep[e.g.,][]{Cabrit90,Hartigan95,Calvet97}.
If an increase in mass accretion rate would induce a rise in the optical continuum flux, it might produce a larger velocity for the ejecta, and therefore causes internal shocks (and therefore `a knot') in the jet as newly ejected gas hits slower gas ejected earlier \citep[e.g.,][]{Raga90,White14a}.

As the \ion{Ca}{2} and \ion{O}{1} emission we observed are associated with the region very close to the star ($\ll$0.1 au; Section \ref{sec:discussion11}), this would also suggest that the jet launching region must be located very close to the star as predicted by the X-wind and the reconnection wind models (Section \ref{sec:intro}). In contrast, the disk wind models, in which the jet launching region covers the disk surface up to a few au scale, would yield a  time delay of $\sim$100 days to a few years for a change in the optical continuum flux and line profiles after a jet knot is ejected, as estimated using the equations below:
\begin{equation}
t_{\mathrm delay} = r/c_s = r (k_B T/\mu)^{-1/2},
\end{equation}
where $r$ is a typical jet launching radius at the disk; $c_s$ is the sound speed; $k_B$ is the Boltzmann constant; $T$ is the temperature at $r$; and $\mu$ is the mean molecular mass. Assuming that the disk surface is heated by stellar radiation, the temperature $T$ would be as follows:
\begin{equation}
T = \left( \frac{L_*}{4 \pi \sigma} \right)^{1/4} r^{-1/2}. 
\end{equation}
where $L_*$ is the stellar luminosity including accretion hotspots; and $\sigma$ is the Stephan-Boltzmann constant. Substituting Equation (2) to Equation (1), we derive:
\begin{equation}
t_{\mathrm delay} = 1.4 \times 10^3 \left( \frac{L_*}{L_\sun} \right)^{-1/8} \left( \frac{r}{\mathrm{1~au}} \right)^{5/4}~\mathrm{days}.
\end{equation}
Adopting $L_*$=1.7 $L_\sun$ (Table \ref{tbl:rwaur}) we would estimate a time delay of $\sim$70,  $\sim$3$\times$10$^2$, and $\sim$1$\times$10$^3$ days for a launching disk radius $r$ of 0.1, 0.3 and 1 au, respectively. Figures \ref{fig:Vmag} and \ref{fig:profs:knots} do not clearly show a time delay of $\gtrsim$3$\times$10$^2$ days for the photometric rises after the ejections of Knots C-E. Our observations may not exclude the possibility of the presence of a time delay of $\lesssim$100 days, for jet launching radii ($r$$\lesssim$0.1 au) which is extremely small for the disk wind models \citep[e.g.,][]{Coffey15}.
{

Figure \ref{fig:Vmag} shows some trends which cannot be simply attributed to the above explanation. First, we do not find a jet knot in Figure \ref{fig:jet} corresponding to a photometric rise in late 2017. As shown in Figure \ref{fig:Vmag}, the extent of this photometric rise is $\Delta V$=1.3 mag., significantly lower than those at/near Knots CDE (1.9--2.6 mag.). Therefore, the velocity increase of the jet induced by a modest accretion rate might not have been sufficient to induce shocks bright enough to be identified in our observations. 
Secondly, Figure \ref{fig:Vmag} does not clearly show evidence for a photometric rise associated with Knot B. One of the following two situations would explain this trend. First, we might have missed a photometric rise in early 2006 or 2007, for which we do not have photometric data. Secondly, Knot B is significantly fainter than the others in Figure \ref{fig:jet}, therefore its origin might be different from the others, 
and it might not be directly related to time variable mass accretion.

Measurements of the proper motion of Knot E for the next few years, and measurements with another new knot ejection, are required to confirm or reject the link between the jet knot ejections and the photometric rises. If confirmed, it would significantly constrain the location of the jet knot ejections, the mechanism of jet ejection and a link with mass accretion, as discussed above. Furthermore, the photometric and spectroscopic variabilities of RW Aur A are exceptionally complicated among pre-main sequence stars, therefore similar studies with another few stars would also be useful for investigating these physical mechanisms, which are essential for star formation. In addition, jet observations at a significantly higher resolution might become possible in future, and these observations would be useful for testing the above scenario of jet knot formation.


\section{Conclusions}  \label{sec:conclusions}
For the active T-Taur star RW Aur A we have performed long-term ($\sim$10 yr) monitoring observations of (1) jet structures in the  [\ion{Fe}{2}] 1.644 \micron~emission using Gemini-NIFS and VLT-SINFONI; (2) optical high-resolution spectroscopy using CFHT-ESPaDOnS; and (3) optical photometry. The latter two observations confirm a correlation of time variabilities between (A) the \ion{Ca}{2} 8542 \AA~ and \ion{O}{1} 7772 \AA~line profiles associated with magnetospheric accretion, and (B) optical continuum fluxes, previously reported by \citet{Takami16} using part of these data sets. The proper motions of jet knots shown in seven epochs of the observations indicate that four knot ejections occurred at the star over the past $\sim$15 years with an irregular interval of 2-6 year. The time scale and irregularity of these intervals are similar to that of the dimming events seen in the optical photometry data since 2010 (1-5 years).

The above observations show a possible link between remarkable ($\Delta V < -1.5$) photometric rise and jet knot ejections. Observations over another few years may confirm or reject this trend. If confirmed, it would imply that the location of the jet launching region is very close to the star ($r \lesssim 0.1$ au) as predicted for some jet launching models. Such a conclusion would be crucial for understanding disk evolution within a few au of the star, and therefore possible ongoing planet formation at these radii.


\acknowledgments
We thank Dr. Elena Valenti for reducing the SINFONI data with the ESO pipeline.
We are grateful to Drs. Chian-Chou Chen and Ming-Yi Lin for useful discussion.
We thank the Gemini Observatory staff for their assistance preparing our programs for data acquisition, and we thank staff observers for executing our program observations during the assigned queue time. 
M.T. is supported by the Ministry of Science and Technology (MoST) of Taiwan (grant No. 106-2119-M-001-026-MY3).
R.G.M. acknowledges support from UNAM-PAPIIT project IN104319.
T.P.R. acknowledges support from the European Research Council through grant No. 743029.
Based on observations obtained at the Gemini Observatory, which is operated by the Association of Universities for Research in Astronomy, Inc., under a cooperative agreement with the NSF on behalf of the Gemini partnership: the National Science Foundation (United States), National Research Council (Canada), CONICYT (Chile), Ministerio de Ciencia, Tecnologa e Innovacin Productiva (Argentina), Ministrio da Cincia, Tecnologia e Inovao (Brazil), and Korea Astronomy and Space Science Institute (Republic of Korea).
This work has made use of data from the European Space Agency (ESA) mission Gaia (https://www.cosmos.esa.int/gaia), processed by the Gaia Data Processing and Analysis Consortium (DPAC, https://www.cosmos.esa.int/web/gaia/dpac/consortium). Funding for the DPAC has been provided by national institutions, in particular the institutions participating in the Gaia Multilateral Agreement.
This research made use of the Simbad database operated at CDS, Strasbourg, France, and the NASA's Astrophysics Data System Abstract Service.

%

\vspace{5mm}
\facilities{Gemini(NIFS), VLT(SINFONI), CFHT(ESPaDOnS), CrAO 1.25m}


\software{
IRAF \citep{Tody86,Tody93},
PyRAF \citep{pyraf},
numpy \citep{numpy},
scipy \citep{scipy},
astropy \citep{astropy}, 
Gemini IRAF package: https://www.gemini.edu/sciops/data-and-results/processingsoftware,
ESO Refrex
          }










\begin{thebibliography}{}
\expandafter\ifx\csname natexlab\endcsname\relax\def\natexlab#1{#1}\fi
\providecommand{\url}[1]{\href{#1}{#1}}
\providecommand{\dodoi}[1]{doi:~\href{http://doi.org/#1}{\nolinkurl{#1}}}
\providecommand{\doeprint}[1]{\href{http://ascl.net/#1}{\nolinkurl{http://ascl.net/#1}}}
\providecommand{\doarXiv}[1]{\href{https://arxiv.org/abs/#1}{\nolinkurl{https://arxiv.org/abs/#1}}}

\bibitem[{{Agra-Amboage} {et~al.}(2011){Agra-Amboage}, {Dougados}, {Cabrit}, \&
  {Reunanen}}]{Agra11}
{Agra-Amboage}, V., {Dougados}, C., {Cabrit}, S., \& {Reunanen}, J. 2011, \aap,
  532, A59, \dodoi{10.1051/0004-6361/201015886}

\bibitem[{{Alencar} {et~al.}(2005){Alencar}, {Basri}, {Hartmann}, \&
  {Calvet}}]{Alencar05}
{Alencar}, S.~H.~P., {Basri}, G., {Hartmann}, L., \& {Calvet}, N. 2005, \aap,
  440, 595, \dodoi{10.1051/0004-6361:20053315}

\bibitem[{{Antipin} {et~al.}(2015){Antipin}, {Belinski}, {Cherepashchuk},
  {Cherjasov}, {Dodin}, {Gorbunov}, {Lamzin}, {Kornilov}, {Kornilov},
  {Potanin}, {Safonov}, {Senik}, {Shatsky}, \& {Voziakova}}]{Antipin15}
{Antipin}, S., {Belinski}, A., {Cherepashchuk}, A., {et~al.} 2015, Information
  Bulletin on Variable Stars, 6126, 1.
\newblock \doarXiv{1412.7661}

\bibitem[{{Appenzeller} \& {Wolf}(1982)}]{Appenzeller82}
{Appenzeller}, I., \& {Wolf}, B. 1982, \aap, 105, 313

\bibitem[{{Astropy Collaboration} {et~al.}(2013){Astropy Collaboration},
  {Robitaille}, {Tollerud}, {Greenfield}, {Droettboom}, {Bray}, {Aldcroft},
  {Davis}, {Ginsburg}, {Price-Whelan}, {Kerzendorf}, {Conley}, {Crighton},
  {Barbary}, {Muna}, {Ferguson}, {Grollier}, {Parikh}, {Nair}, {Unther},
  {Deil}, {Woillez}, {Conseil}, {Kramer}, {Turner}, {Singer}, {Fox}, {Weaver},
  {Zabalza}, {Edwards}, {Azalee Bostroem}, {Burke}, {Casey}, {Crawford},
  {Dencheva}, {Ely}, {Jenness}, {Labrie}, {Lim}, {Pierfederici}, {Pontzen},
  {Ptak}, {Refsdal}, {Servillat}, \& {Streicher}}]{astropy}
{Astropy Collaboration}, {Robitaille}, T.~P., {Tollerud}, E.~J., {et~al.} 2013,
  \aap, 558, A33, \dodoi{10.1051/0004-6361/201322068}

\bibitem[{{Bacciotti} {et~al.}(1996){Bacciotti}, {Hirth}, \&
  {Natta}}]{Bacciotti96}
{Bacciotti}, F., {Hirth}, G.~A., \& {Natta}, A. 1996, \aap, 310, 309

\bibitem[{{Bacciotti} {et~al.}(2002){Bacciotti}, {Ray}, {Mundt},
  {Eisl{\"o}ffel}, \& {Solf}}]{Bacciotti02}
{Bacciotti}, F., {Ray}, T.~P., {Mundt}, R., {Eisl{\"o}ffel}, J., \& {Solf}, J.
  2002, \apj, 576, 222, \dodoi{10.1086/341725}

\bibitem[{{Beck} {et~al.}(2008){Beck}, {McGregor}, {Takami}, \& {Pyo}}]{Beck08}
{Beck}, T.~L., {McGregor}, P.~J., {Takami}, M., \& {Pyo}, T. 2008, \apj, 676,
  472, \dodoi{10.1086/527528}

\bibitem[{{Beck} \& {Simon}(2001)}]{Beck01}
{Beck}, T.~L., \& {Simon}, M. 2001, \aj, 122, 413, \dodoi{10.1086/321133}

\bibitem[{{Berdnikov} {et~al.}(2017){Berdnikov}, {Burlak}, {Vozyakova},
  {Dodin}, {Lamzin}, \& {Tatarnikov}}]{Berdnikov17}
{Berdnikov}, L.~N., {Burlak}, M.~A., {Vozyakova}, O.~V., {et~al.} 2017,
  Astrophysical Bulletin, 72, 277, \dodoi{10.1134/S1990341317030178}

\bibitem[{{Bisikalo} {et~al.}(2012){Bisikalo}, {Dodin}, {Kaigorodov}, {Lamzin},
  {Malogolovets}, \& {Fateeva}}]{Bisikalo12}
{Bisikalo}, D.~V., {Dodin}, A.~V., {Kaigorodov}, P.~V., {et~al.} 2012,
  Astronomy Reports, 56, 686, \dodoi{10.1134/S1063772912090028}

\bibitem[{{Blandford} \& {Payne}(1982)}]{Blandford82}
{Blandford}, R.~D., \& {Payne}, D.~G. 1982, \mnras, 199, 883

\bibitem[{{Bouvier} {et~al.}(2007){Bouvier}, {Alencar}, {Harries},
  {Johns-Krull}, \& {Romanova}}]{Bouvier07_PPV}
{Bouvier}, J., {Alencar}, S.~H.~P., {Harries}, T.~J., {Johns-Krull}, C.~M., \&
  {Romanova}, M.~M. 2007, Protostars and Planets V, 479

\bibitem[{{Bouvier} {et~al.}(2014){Bouvier}, {Matt}, {Mohanty}, {Scholz},
  {Stassun}, \& {Zanni}}]{Bouvier14}
{Bouvier}, J., {Matt}, S.~P., {Mohanty}, S., {et~al.} 2014, Protostars and
  Planets VI, 433, \dodoi{10.2458/azu_uapress_9780816531240-ch019}

\bibitem[{{Bozhinova} {et~al.}(2016){Bozhinova}, {Scholz}, {Costigan}, {Lux},
  {Davis}, {Ray}, {Boardman}, {Hay}, {Hewlett}, {Hodos{\'a}n}, \&
  {Morton}}]{Bozhinova16}
{Bozhinova}, I., {Scholz}, A., {Costigan}, G., {et~al.} 2016, \mnras, 463,
  4459, \dodoi{10.1093/mnras/stw2327}

\bibitem[{{Cabrit} {et~al.}(1990){Cabrit}, {Edwards}, {Strom}, \&
  {Strom}}]{Cabrit90}
{Cabrit}, S., {Edwards}, S., {Strom}, S.~E., \& {Strom}, K.~M. 1990, \apj, 354,
  687, \dodoi{10.1086/168725}

\bibitem[{{Calvet}(1997)}]{Calvet97}
{Calvet}, N. 1997, in IAU Symposium, Vol. 182, Herbig-Haro Flows and the Birth
  of Stars, ed. B.~{Reipurth} \& C.~{Bertout}, 417--432

\bibitem[{{Calvet} {et~al.}(2000){Calvet}, {Hartmann}, \& {Strom}}]{Calvet00}
{Calvet}, N., {Hartmann}, L., \& {Strom}, S.~E. 2000, Protostars and Planets
  IV, 377

\bibitem[{{Chou} {et~al.}(2013){Chou}, {Takami}, {Manset}, {Beck}, {Pyo},
  {Chen}, {Panwar}, {Karr}, {Shang}, \& {Liu}}]{Chou13}
{Chou}, M.-Y., {Takami}, M., {Manset}, N., {et~al.} 2013, \aj, 145, 108,
  \dodoi{10.1088/0004-6256/145/4/108}

\bibitem[{{Coffey} {et~al.}(2008){Coffey}, {Bacciotti}, \& {Podio}}]{Coffey08}
{Coffey}, D., {Bacciotti}, F., \& {Podio}, L. 2008, \apj, 689, 1112,
  \dodoi{10.1086/592343}

\bibitem[{{Coffey} {et~al.}(2004){Coffey}, {Bacciotti}, {Woitas}, {Ray}, \&
  {Eisl{\"o}ffel}}]{Coffey04}
{Coffey}, D., {Bacciotti}, F., {Woitas}, J., {Ray}, T.~P., \& {Eisl{\"o}ffel},
  J. 2004, \apj, 604, 758, \dodoi{10.1086/382019}

\bibitem[{{Coffey} {et~al.}(2015){Coffey}, {Dougados}, {Cabrit}, {Pety}, \&
  {Bacciotti}}]{Coffey15}
{Coffey}, D., {Dougados}, C., {Cabrit}, S., {Pety}, J., \& {Bacciotti}, F.
  2015, \apj, 804, 2, \dodoi{10.1088/0004-637X/804/1/2}

\bibitem[{{Coffey} {et~al.}(2012){Coffey}, {Rigliaco}, {Bacciotti}, {Ray}, \&
  {Eisl{\"o}ffel}}]{Coffey12}
{Coffey}, D., {Rigliaco}, E., {Bacciotti}, F., {Ray}, T.~P., \&
  {Eisl{\"o}ffel}, J. 2012, \apj, 749, 139, \dodoi{10.1088/0004-637X/749/2/139}

\bibitem[{{Dodin} {et~al.}(2019){Dodin}, {Grankin}, {Lamzin}, {Nadjip},
  {Safonov}, {Shakhovskoi}, {Shenavrin}, {Tatarnikov}, \&
  {Vozyakova}}]{Dodin19}
{Dodin}, A., {Grankin}, K., {Lamzin}, S., {et~al.} 2019, \mnras, 482, 5524,
  \dodoi{10.1093/mnras/sty2988}

\bibitem[{{Dougados} {et~al.}(2000){Dougados}, {Cabrit}, {Lavalley}, \&
  {M{\'e}nard}}]{Dougados00}
{Dougados}, C., {Cabrit}, S., {Lavalley}, C., \& {M{\'e}nard}, F. 2000, \aap,
  357, L61

\bibitem[{{Facchini} {et~al.}(2016){Facchini}, {Manara}, {Schneider}, {Clarke},
  {Bouvier}, {Rosotti}, {Booth}, \& {Haworth}}]{Facchini16}
{Facchini}, S., {Manara}, C.~F., {Schneider}, P.~C., {et~al.} 2016, \aap, 596,
  A38, \dodoi{10.1051/0004-6361/201629607}

\bibitem[{{Frank} {et~al.}(2014){Frank}, {Ray}, {Cabrit}, {Hartigan}, {Arce},
  {Bacciotti}, {Bally}, {Benisty}, {Eisl{\"o}ffel}, {G{\"u}del}, {Lebedev},
  {Nisini}, \& {Raga}}]{Frank14}
{Frank}, A., {Ray}, T.~P., {Cabrit}, S., {et~al.} 2014, Protostars and Planets
  VI, 451, \dodoi{10.2458/azu_uapress_9780816531240-ch020}

\bibitem[{{Gahm}(1970)}]{Gahm70}
{Gahm}, G.~F. 1970, \apj, 160, 1117, \dodoi{10.1086/150498}

\bibitem[{{Gahm} {et~al.}(1999){Gahm}, {Petrov}, {Duemmler}, {Gameiro}, \&
  {Lago}}]{Gahm99}
{Gahm}, G.~F., {Petrov}, P.~P., {Duemmler}, R., {Gameiro}, J.~F., \& {Lago},
  M.~T.~V.~T. 1999, \aap, 352, L95

\bibitem[{{Gaia Collaboration} {et~al.}(2018){Gaia Collaboration}, {Brown},
  {Vallenari}, {Prusti}, {de Bruijne}, {Babusiaux}, {Bailer-Jones}, {Biermann},
  {Evans}, {Eyer}, \& et~al.}]{GaiaDR2}
{Gaia Collaboration}, {Brown}, A.~G.~A., {Vallenari}, A., {et~al.} 2018, \aap,
  616, A1, \dodoi{10.1051/0004-6361/201833051}

\bibitem[{{G{\'a}rate} {et~al.}(2019){G{\'a}rate}, {Birnstiel}, {Stammler}, \&
  {G{\"u}nther}}]{Garate18}
{G{\'a}rate}, M., {Birnstiel}, T., {Stammler}, S.~M., \& {G{\"u}nther}, H.~M.
  2019, \apj, 871, 53, \dodoi{10.3847/1538-4357/aaf4fc}

\bibitem[{{Grankin} {et~al.}(2007){Grankin}, {Melnikov}, {Bouvier}, {Herbst},
  \& {Shevchenko}}]{Grankin07}
{Grankin}, K.~N., {Melnikov}, S.~Y., {Bouvier}, J., {Herbst}, W., \&
  {Shevchenko}, V.~S. 2007, \aap, 461, 183, \dodoi{10.1051/0004-6361:20065489}

\bibitem[{{Gullbring} {et~al.}(1998){Gullbring}, {Hartmann}, {Brice{\~n}o}, \&
  {Calvet}}]{Gullbring98}
{Gullbring}, E., {Hartmann}, L., {Brice{\~n}o}, C., \& {Calvet}, N. 1998, \apj,
  492, 323, \dodoi{10.1086/305032}

\bibitem[{{G{\"u}nther} {et~al.}(2018){G{\"u}nther}, {Birnstiel},
  {Huenemoerder}, {Principe}, {Schneider}, {Wolk}, {Dubois}, {Logie}, {Rau}, \&
  {Vanaverbeke}}]{Gunther18}
{G{\"u}nther}, H.~M., {Birnstiel}, T., {Huenemoerder}, D.~P., {et~al.} 2018,
  \aj, 156, 56, \dodoi{10.3847/1538-3881/aac9bd}

\bibitem[{{Hartigan} {et~al.}(1995){Hartigan}, {Edwards}, \&
  {Ghandour}}]{Hartigan95}
{Hartigan}, P., {Edwards}, S., \& {Ghandour}, L. 1995, \apj, 452, 736,
  \dodoi{10.1086/176344}

\bibitem[{{Hartigan} \& {Hillenbrand}(2009)}]{Hartigan09}
{Hartigan}, P., \& {Hillenbrand}, L. 2009, \apj, 705, 1388,
  \dodoi{10.1088/0004-637X/705/2/1388}

\bibitem[{{Herbig}(1945)}]{Herbig45}
{Herbig}, G.~H. 1945, \pasp, 57, 166, \dodoi{10.1086/125709}

\bibitem[{{Herbig}(1948)}]{Herbig48}
---. 1948, \pasp, 60, 256, \dodoi{10.1086/126057}

\bibitem[{{Herbst} {et~al.}(1994){Herbst}, {Herbst}, {Grossman}, \&
  {Weinstein}}]{Herbst94}
{Herbst}, W., {Herbst}, D.~K., {Grossman}, E.~J., \& {Weinstein}, D. 1994, \aj,
  108, 1906, \dodoi{10.1086/117204}

\bibitem[{{Hirth} {et~al.}(1997){Hirth}, {Mundt}, \& {Solf}}]{Hirth97}
{Hirth}, G.~A., {Mundt}, R., \& {Solf}, J. 1997, \aaps, 126, 437,
  \dodoi{10.1051/aas:1997275}

\bibitem[{{Hirth} {et~al.}(1994){Hirth}, {Mundt}, {Solf}, \& {Ray}}]{Hirth94}
{Hirth}, G.~A., {Mundt}, R., {Solf}, J., \& {Ray}, T.~P. 1994, \apjl, 427, L99,
  \dodoi{10.1086/187374}

\bibitem[{Jones {et~al.}(2001)Jones, Oliphant, Peterson, {et~al.}}]{scipy}
Jones, E., Oliphant, T., Peterson, P., {et~al.} 2001, {SciPy}: Open source
  scientific tools for {Python}.
\newblock \url{http://www.scipy.org/}

\bibitem[{{Joy} \& {van Biesbroeck}(1944)}]{Joy44}
{Joy}, A.~H., \& {van Biesbroeck}, G. 1944, \pasp, 56, 123,
  \dodoi{10.1086/125628}

\bibitem[{{Kafka}(2018)}]{aavso}
{Kafka}, S. 2018, {Observations from the AAVSO International Database}.
\newblock \url{https://www.aavso.org/}

\bibitem[{{K\"onigl} \& {Pudritz}(2000)}]{Konigl00}
{K\"onigl}, A., \& {Pudritz}, R.~E. 2000, Protostars and Planets IV, 759

\bibitem[{{Koutoulaki} {et~al.}(2019){Koutoulaki}, {Facchini}, {Manara},
  {Natta}, {Garcia Lopez}, {Fedriani}, {Caratti o Garatti}, {Coffey}, \&
  {Ray}}]{Koutoulaki19}
{Koutoulaki}, M., {Facchini}, S., {Manara}, C.~F., {et~al.} 2019, \aap, 625,
  A49, \dodoi{10.1051/0004-6361/201834713}

\bibitem[{{Kurosawa} \& {Romanova}(2013)}]{Kurosawa13}
{Kurosawa}, R., \& {Romanova}, M.~M. 2013, \mnras, 431, 2673,
  \dodoi{10.1093/mnras/stt365}

\bibitem[{{Lamzin} {et~al.}(2017){Lamzin}, {Cheryasov}, {Chuntonov}, {Dodin},
  {Grankin}, {Malanchev}, {Nadzhip}, {Safonov}, {Shakhovskoy}, {Shenavrin},
  {Tatarnikov}, \& {Vozyakova}}]{Lamzin17}
{Lamzin}, S., {Cheryasov}, D., {Chuntonov}, G., {et~al.} 2017, in Astronomical
  Society of the Pacific Conference Series, Vol. 510, Stars: From Collapse to
  Collapse, ed. Y.~Y. {Balega}, D.~O. {Kudryavtsev}, I.~I. {Romanyuk}, \& I.~A.
  {Yakunin}, 356.
\newblock \doarXiv{1707.09671}

\bibitem[{{Lamzin}(1999)}]{Lamzin99}
{Lamzin}, S.~A. 1999, Astronomy Letters, 25, 430

\bibitem[{{Lee} {et~al.}(2017){Lee}, {Ho}, {Li}, {Hirano}, {Zhang}, \&
  {Shang}}]{Lee17}
{Lee}, C.-F., {Ho}, P. T.~P., {Li}, Z.-Y., {et~al.} 2017, Nature Astronomy, 1,
  0152, \dodoi{10.1038/s41550-017-0152}

\bibitem[{{Liu} \& {Shang}(2012)}]{Liu12}
{Liu}, C.-F., \& {Shang}, H. 2012, \apj, 761, 94,
  \dodoi{10.1088/0004-637X/761/2/94}

\bibitem[{{L{\'o}pez-Mart{\'{\i}}n} {et~al.}(2003){L{\'o}pez-Mart{\'{\i}}n},
  {Cabrit}, \& {Dougados}}]{Lopez03}
{L{\'o}pez-Mart{\'{\i}}n}, L., {Cabrit}, S., \& {Dougados}, C. 2003, \aap, 405,
  L1, \dodoi{10.1051/0004-6361:20030758}

\bibitem[{{Lord}(1992)}]{ATRAN}
{Lord}, S.~D. 1992, {A new software tool for computing Earth's atmospheric
  transmission of near- and far-infrared radiation}, NASA Technical Memorandum
  103957

\bibitem[{{Machida} {et~al.}(2008){Machida}, {Inutsuka}, \&
  {Matsumoto}}]{Machida08}
{Machida}, M.~N., {Inutsuka}, S., \& {Matsumoto}, T. 2008, \apj, 676, 1088,
  \dodoi{10.1086/528364}

\bibitem[{{Melnikov} {et~al.}(2009){Melnikov}, {Eisl{\"o}ffel}, {Bacciotti},
  {Woitas}, \& {Ray}}]{Melnikov09}
{Melnikov}, S.~Y., {Eisl{\"o}ffel}, J., {Bacciotti}, F., {Woitas}, J., \&
  {Ray}, T.~P. 2009, \aap, 506, 763, \dodoi{10.1051/0004-6361/200811567}

\bibitem[{{Mundt} \& {Eisl{\"o}ffel}(1998)}]{Mundt98}
{Mundt}, R., \& {Eisl{\"o}ffel}, J. 1998, \aj, 116, 860, \dodoi{10.1086/300461}

\bibitem[{{Najita} {et~al.}(2000){Najita}, {Edwards}, {Basri}, \&
  {Carr}}]{Najita00}
{Najita}, J.~R., {Edwards}, S., {Basri}, G., \& {Carr}, J. 2000, Protostars and
  Planets IV, 457

\bibitem[{{Oliphant}(2006)}]{numpy}
{Oliphant}, T.~E. 2006, A Guide to NumPy (Trelgol Publishing, USA)

\bibitem[{{Petrov} {et~al.}(2015){Petrov}, {Gahm}, {Djupvik}, {Babina},
  {Artemenko}, \& {Grankin}}]{Petrov15}
{Petrov}, P.~P., {Gahm}, G.~F., {Djupvik}, A.~A., {et~al.} 2015, \aap, 577,
  A73, \dodoi{10.1051/0004-6361/201525845}

\bibitem[{{Petrov} {et~al.}(2001){Petrov}, {Gahm}, {Gameiro}, {Duemmler},
  {Ilyin}, {Laakkonen}, {Lago}, \& {Tuominen}}]{Petrov01a}
{Petrov}, P.~P., {Gahm}, G.~F., {Gameiro}, J.~F., {et~al.} 2001, \aap, 369,
  993, \dodoi{10.1051/0004-6361:20010203}

\bibitem[{{Pudritz} \& {Norman}(1983)}]{Pudritz83}
{Pudritz}, R.~E., \& {Norman}, C.~A. 1983, \apj, 274, 677,
  \dodoi{10.1086/161481}

\bibitem[{{Pyo} {et~al.}(2003){Pyo}, {Kobayashi}, {Hayashi}, {Terada}, {Goto},
  {Takami}, {Takato}, {Gaessler}, {Usuda}, {Yamashita}, {Tokunaga}, {Hayano},
  {Kamata}, {Iye}, \& {Minowa}}]{Pyo03}
{Pyo}, T.-S., {Kobayashi}, N., {Hayashi}, M., {et~al.} 2003, \apj, 590, 340,
  \dodoi{10.1086/374966}

\bibitem[{{Pyo} {et~al.}(2006){Pyo}, {Hayashi}, {Kobayashi}, {Tokunaga},
  {Terada}, {Takami}, {Takato}, {Davis}, {Takami}, {Hayashi}, {G{\"a}ssler},
  {Oya}, {Hayano}, {Kamata}, {Minowa}, {Iye}, {Usuda}, {Nishikawa}, \&
  {Nedachi}}]{Pyo06}
{Pyo}, T.-S., {Hayashi}, M., {Kobayashi}, N., {et~al.} 2006, \apj, 649, 836,
  \dodoi{10.1086/506929}

\bibitem[{{Raga} {et~al.}(1990){Raga}, {Canto}, {Binette}, \&
  {Calvet}}]{Raga90}
{Raga}, A.~C., {Canto}, J., {Binette}, L., \& {Calvet}, N. 1990, \apj, 364,
  601, \dodoi{10.1086/169443}

\bibitem[{{Reipurth} \& {Zinnecker}(1993)}]{Reipurth93}
{Reipurth}, B., \& {Zinnecker}, H. 1993, \aap, 278, 81

\bibitem[{{Rodriguez} {et~al.}(2013){Rodriguez}, {Pepper}, {Stassun}, {Siverd},
  {Cargile}, {Beatty}, \& {Gaudi}}]{Rodriguez13}
{Rodriguez}, J.~E., {Pepper}, J., {Stassun}, K.~G., {et~al.} 2013, \aj, 146,
  112, \dodoi{10.1088/0004-6256/146/5/112}

\bibitem[{{Rodriguez} {et~al.}(2018){Rodriguez}, {Loomis}, {Cabrit}, {Haworth},
  {Facchini}, {Dougados}, {Booth}, {Jensen}, {Clarke}, {Stassun}, {Dent}, \&
  {Pety}}]{Rodriguez18}
{Rodriguez}, J.~E., {Loomis}, R., {Cabrit}, S., {et~al.} 2018, \apj, 859, 150,
  \dodoi{10.3847/1538-4357/aac08f}

\bibitem[{{Romanova} {et~al.}(2008){Romanova}, {Kulkarni}, \&
  {Lovelace}}]{Romanova08}
{Romanova}, M.~M., {Kulkarni}, A.~K., \& {Lovelace}, R.~V.~E. 2008, \apjl, 673,
  L171, \dodoi{10.1086/527298}

\bibitem[{{Schneider} {et~al.}(2015){Schneider}, {G{\"u}nther}, {Robrade},
  {Facchini}, {Hodapp}, {Manara}, {Perdelwitz}, {Schmitt}, {Skinner}, \&
  {Wolk}}]{Schneider15}
{Schneider}, P.~C., {G{\"u}nther}, H.~M., {Robrade}, J., {et~al.} 2015, \aap,
  584, L9, \dodoi{10.1051/0004-6361/201527237}

\bibitem[{{Science Software Branch at STScI}(2012)}]{pyraf}
{Science Software Branch at STScI}. 2012, {PyRAF: Python alternative for IRAF},
  Astrophysics Source Code Library.
\newblock \doeprint{1207.011}

\bibitem[{{Shenavrin} {et~al.}(2015){Shenavrin}, {Petrov}, \&
  {Grankin}}]{Shenavrin15}
{Shenavrin}, V.~I., {Petrov}, P.~P., \& {Grankin}, K.~N. 2015, Information
  Bulletin on Variable Stars, 6143, 1

\bibitem[{{Shu} {et~al.}(2000){Shu}, {Najita}, {Shang}, \& {Li}}]{Shu00}
{Shu}, F.~H., {Najita}, J.~R., {Shang}, H., \& {Li}, Z. 2000, Protostars and
  Planets IV, 789

\bibitem[{{Takami} {et~al.}(2016){Takami}, {Wei}, {Chou}, {Karr}, {Beck},
  {Manset}, {Chen}, {Kurosawa}, {Fukagawa}, {White}, {Galv{\'a}n-Madrid},
  {Liu}, {Pyo}, \& {Donati}}]{Takami16}
{Takami}, M., {Wei}, Y.-J., {Chou}, M.-Y., {et~al.} 2016, \apj, 820, 139,
  \dodoi{10.3847/0004-637X/820/2/139}

\bibitem[{{Tody}(1986)}]{Tody86}
{Tody}, D. 1986, in \procspie, Vol. 627, Instrumentation in astronomy VI, ed.
  D.~L. {Crawford}, 733, \dodoi{10.1117/12.968154}

\bibitem[{{Tody}(1993)}]{Tody93}
{Tody}, D. 1993, in Astronomical Society of the Pacific Conference Series,
  Vol.~52, Astronomical Data Analysis Software and Systems II, ed. R.~J.
  {Hanisch}, R.~J.~V. {Brissenden}, \& J.~{Barnes}, 173

\bibitem[{{White} {et~al.}(2014){White}, {McGregor}, {Bicknell}, {Salmeron}, \&
  {Beck}}]{White14a}
{White}, M.~C., {McGregor}, P.~J., {Bicknell}, G.~V., {Salmeron}, R., \&
  {Beck}, T.~L. 2014, \mnras, 441, 1681, \dodoi{10.1093/mnras/stu654}

\bibitem[{{White} \& {Ghez}(2001)}]{White01}
{White}, R.~J., \& {Ghez}, A.~M. 2001, \apj, 556, 265, \dodoi{10.1086/321542}

\bibitem[{{Woitas} {et~al.}(2002){Woitas}, {Ray}, {Bacciotti}, {Davis}, \&
  {Eisl{\"o}ffel}}]{Woitas02}
{Woitas}, J., {Ray}, T.~P., {Bacciotti}, F., {Davis}, C.~J., \&
  {Eisl{\"o}ffel}, J. 2002, \apj, 580, 336, \dodoi{10.1086/343124}

\end{thebibliography}
\bibliographystyle{aasjournal}



\end{document}